\documentstyle[12pt,preprint,tighten,floats,aps,epsf,epsfig,psfig]{revtex}
\begin{document}
\preprint{IC/99/161}
\newcommand{\newc}{\newcommand}
\def\be{\begin{equation}}
\def\ee{\end{equation}}
\def\bea{\begin{eqnarray}}
\def\eea{\end{eqnarray}}
\def\simlt{\stackrel{<}{{}_\sim}}
\def\simgt{\stackrel{>}{{}_\sim}}
\title{Dynamical relaxation of the CP phases in next--to--minimal supersymmetry}
\author{D. A. Demir}
\address{The Abdus Salam International Centre for Theoretical Physics, I-34100 Trieste, Italy}
\date{\today}
\maketitle
\begin{abstract}
After promoting the phases of the soft masses to dynamical fields corresponding to 
Goldstone bosons of spontaneously broken global symmetries in the supersymmetry breaking 
sector, the next--to--minimal supersymmetric model is found to solve the 
$\mu$ problem and the strong CP problem simultaneously with an invisible axion. 
The domain wall problem persists in the form of axionic domain formation. 
Relaxation dynamics of the physical CP--violating phases is determined only by 
the short--distance physics and their relaxation values are not necessarily close
to the CP--conserving points. Having observable supersymmetric CP violation and
avoiding the axionic domain walls both require nonminimal flavor structures.
\end{abstract}
\section{Introduction}
The present bounds on the electric dipole moments (EDM) of the particles
generate serious hierarchy problems concerning the amount of CP violation. 
To be specific, let us consider the neutron EDM \cite{nedm}. It is well known that the QCD
vacuum angle ($\theta_{QCD}$) induces a neutron EDM which is approximately ten orders of magnitude larger
than the existing bound. This is the source of the so-called strong CP
problem, that is, $|\theta_{QCD}| \simlt 10^{-10}$
instead of the expected order of unity. This naturalness problem
is nicely solved by the celebrated Peccei--Quinn mechanism \cite{pq} 
which promotes $\theta_{QCD}$  to a dynamical variable via the phases
of the quarks \cite{pq,ww,dfsz} or additional color triplets \cite{ksvz},
and then relax it to zero thanks to the instanton--induced effective potential.

In the realm of supersymmetry in particular the minimal supersymmetric
standard model (MSSM), there arises SUSY CP problem in addition to 
the strong CP problem. Indeed, the soft--breaking parameters as well as
the $\mu$ parameter can have nonvanishing phases \cite{ben} leading to
a neutron EDM exceeding the bounds by three orders of magnitude, 
except for certain portions of the SUSY parameter space where different
sparticle contributions cancel \cite{cancel}. However, it is clear that 
even if SUSY contributions partially cancel to agree with the 
experiment this is by no means a complete solution of the problem because 
the strong CP problem is still there.
 
A simultaneous solution to both strong CP and SUSY CP problems
have been shown to exist \cite{relaxMSSM} if there are spontaneously
broken global symmetries in the SUSY--breaking sector. Here 
all soft masses as well as the $\mu$ parameter possess  
dynamical phases which realize the global symmetries in 
the SUSY--breaking sector nonlinearly. In effect, the relaxation
mechanism proposed in \cite{relaxMSSM} involves promoting
the phases of the soft masses to dynamical variables so that 
($i$) the phases in the quark and gluino mass matrices  
relax the QCD vacuum angle, and ($ii$) all phases
appearing in the vacuum energy relax to CP--conserving points
due to the radiative stability of the vacuum energy. While former solves
the strong CP problem the latter does the SUSY CP problem. Therefore,
the end result of this mechanism is that there is no source of
CP violation beyond the Kobayashi--Maskawa (CKM) phase.

However, the MSSM suffers from a serious hierarchy problem: the $\mu$ puzzle.
Namely, the Higgsino bilinear mass parameter, $\mu$ which follows 
from the superpotential, can be anywhere between the weak scale
and the Planck scale \cite{muprob}. In fact, the relaxation mechanism in 
\cite{relaxMSSM} treats the $\mu$ parameter as a soft mass which is already stabilized
at the weak scale. In this context, the next--to--minimal supersymmetric
SM (NMSSM) \cite{nmssm} is the most economic extension of the MSSM in which the 
$\mu$ parameter is induced by the VEV of an additional gauge singlet. One notes
that the NMSSM not only solves the $\mu$ problem but also offers a rich
phenomenology for colliders \cite{pheno} and dark matter \cite{dark}.

In this work we will discuss dynamical relaxation of the CP
phases in the NMSSM in order to check if it is possible to have 
a simultaneous solution to the hierarchy problems concerning the 
strong CP, SUSY CP and the $\mu$ parameter.
It will be seen that the generalization of the $\mu$ parameter to
a local operator modifies the infrared and ultraviolet sensitivities of 
the vacuum energy whereby offering a different relaxation mechanism. In particular, 
the radiative stability of the vacuum energy will no longer be sufficient 
to relax all CP--violating phases to CP--conserving points. 
In fact, there will be remnant physical phases that can contribute 
to CP--violating quantities like EDMs or neutral meson mixings.

In Sec. II we identify the possible sources of explicit CP violation
in the tree level NMSSM lagrangian. We then determine the 
symmetries of the superpotential, and thereby show the need for
promoting the phases to dynamical variables. Here we also list down 
all possible phase--dependent operators which can contribute to the
vacuum energy.

In Sec. III we analyze the dynamical relaxation processes for the 
phases. First we discuss the relaxation of the QCD vacuum angle. 
Next we estimate radiative corrections to the vacuum energy, and 
show that the SUSY CP--violating phases no longer relax to the 
CP--conserving points. 

In Sec. IV we discuss the MSSM limit and show that the differences 
between the two models follow mainly from their global symmetries
and assumptions about  the $\mu$ parameter.

In Sec. V we summarize the main results of the work and 
compare them with those of the MSSM in a tabular manner.
We discuss also the implications of the results for 
other CP--violation phenomena.

\section{CP--violating phases in the NMSSM}
The superpotential of the NMSSM is given by
\begin{eqnarray}
\label{nmssm}   
\hat{W}=h_{s}\hat{S} \hat{H}_{u}\cdot \hat{H}_{d} + \frac{1}{3}k_{s} \hat{S}^{3} + 
h_{u} \hat{Q}\cdot \hat{H}_{u} \hat{U}^{c} +
h_{d}\hat{Q}\cdot\hat{H}_{d} \hat{D}^{c}+ 
h_{e} \hat{L}\cdot \hat{H}_{d} \hat{E}^{c}\,
\end{eqnarray}
where the Yukawa couplings $h_s,\cdots,h_e$ are non--hermitian matrices in the
flavor space. The lagrangian of the model consists of several complex parameters
\begin{eqnarray}
\label{pot} 
-{\cal{L}}&\ni&\Bigg( h_{s}k_{s}^{\ast} {S^{\ast}}^{2}  H_{u}\cdot H_{d} +
\mbox{H.c.}
\Bigg)+\Bigg(\frac{1}{2} m_{\lambda}\tilde{\lambda}\tilde{\lambda}
+A_{s}S H_{u}\cdot
H_{d}+\frac{1}{3}A_{k}S^{3}\nonumber\\&+&
A_{u}Q\cdot H_{u} U^{c} + A_{d}Q\cdot
H_{d} D^{c}+ A_{e}L\cdot H_{d} E^{c} + \mbox{H.c.} \Bigg)
\end{eqnarray}
where the first term is the $F$--term contribution, and the rest are all soft SUSY--breaking ones.
In general the gaugino masses $m_\lambda$ as well as the triscalar couplings $A_{s,\cdots,e}$ are all
complex quantities. The Higgs sector of the theory has three parameters, $h_s k_s^{\ast}$, $A_s$ and
$A_k$ which can violate CP. After all phase redefinitions of the Higgs fields there remains one 
independent phase, say $Arg\{h_s k_s^{\ast}\}$, which violates CP explicitly. This can be seen from 
the fact that the mass--eigenstate scalars are mixtures of different CP eigenstates \cite{japan}. 
This tree level CP violation is a property of the NMSSM. For comparison, one notes that in the MSSM
there is no such effect because possible phase of $m_{12}^{2}$ appearing in $m_{12}^{2} H_u\cdot H_d$
can be rotated away after a phase redefinition of the Higgs doublets. Therefore, the only way for generating
CP violation in the MSSM comes by the radiative corrections \cite{ben}.

The NMSSM superpotential (\ref{nmssm}) possesses a global continious $R$--symmetry, ${\cal{U}}(1)_{R}$ \footnote{
Under an $R$--rotation $\theta$ variable has charge $R_{\theta}=R_{\lambda}=R_{\hat{W}}/2$. If a chiral superfield 
has charge $R_{\chi}$ then the charges of its scalar and fermionic components are, respectively, $R_{\chi}$ and
$R_{\chi}-R_{\theta}$.}. In fact, this would be a symmetry of the whole lagrangian were it not for the finite gaugino masses 
($m_{\lambda}$) and $A$--terms. Therefore, the soft--breaking terms break the  ${\cal{U}}(1)_{R}$ 
symmetry down to its $Z_3$ subgroup which causes the formation of domain walls \cite{tamvakis}.
With finite soft masses the only way of keeping ${\cal{U}}(1)_{R}$ symmetry unbroken is to let them have dynamical phases
transforming as
\begin{eqnarray}
\label{rchar}
\left( m_{\lambda}, A_{s}, A_{k}, A_{u}, A_{d}, A_{e}\right) \rightarrow e^{-i R_{\hat{W}}\alpha} 
\left( m_{\lambda}, A_{s}, A_{k}, A_{u}, A_{d}, A_{e}\right)
\end{eqnarray}
where $\alpha$ is the ${\cal{U}}(1)_{R}$ rotation angle, and $R_{\hat{W}}$ is the charge of the 
superpotential (\ref{nmssm}) \cite{Rsym}. Therefore all of the gaugino masses and triscalar couplings
are now spurions with charges compansating those of the fields, and the theory has a unique
global symmetry ${\cal{U}}(1)_{R}$. The same operation of promoting the soft phases to dynamical 
variables in the MSSM, however, results in two global symmetries: an $R$--symmetry and a Peccei--Quinn
symmetry \cite{relaxMSSM}. 

Let us first discuss the tree level vacuum energy. After introducing the dynamical phases (\ref{rchar})
the NMSSM vacuum manifold is described by the Higgs fields, $S$, $H_{u}$ and $H_{d}$, and dynamical phases of 
$A_{s}$ and $A_{k}$. Therefore, a direct minimization of the vacuum energy gives
the relations
\begin{eqnarray}
\label{phaserel}
\mbox{Arg}[S H_{u}\cdot H_{d}]=\mbox{Arg}[A_{s}^\ast],~
\mbox{Arg}[S^{3}]=\mbox{Arg}[A_{k}^\ast],~ \mbox{Arg}[A_{s} A_{k}^{\ast}]=\theta_{s}-\theta_{k}~,
\end{eqnarray}
where $\theta_{s}=\mbox{Arg}[h_s]$, $\theta_{k}=\mbox{Arg}[k_s]$, and 
the Higgs fields here are replaced by their vacuum expectation values (VEV).
These relations fix the phases of the Higgs fields in terms of $A_s$ and $A_k$ phases such that the
relative phase between $A_s$ and $A_k$ equals the phase content of the $F$--term contribution
in (\ref{pot}). Similar relations are also found in the MSSM: $\mbox{Arg}[m_{12}^{2}]=-
\mbox{Arg}[H_{u}\cdot H_{d}]$ after minimizing the vacuum energy \cite{relaxMSSM,ben}.
  
Any physical quantity with a nontrivial phase content is restricted to depend only 
${\cal{U}}(1)_{R}$--invariant combinations of the mass parameters. Since the Higgs
fields eventually acquire VEVs they can also be included
in the list of phase--dependent invariants. Depicted in Table I are sets of
phase--dependent invariants having mass dimension $d=2$ (first column) and 
$d=4$ (second column). The phase of each combination is listed in the third
column in terms of the independent phases 
\begin{eqnarray} 
\label{physphase}
\phi_{s}(x)=\mbox{Arg}[m_{\lambda}A_{s}^{*}],\ \ \
\phi_{k}(x)=\mbox{Arg}[m_{\lambda}A_{k}^{*}],\ \ \ \phi_{f}(x)=\mbox{Arg}[m_{\lambda}A_{f}^{*}]~,
\end{eqnarray}
which  will prove useful in comparing the results with those of the MSSM. 
In Table I, $|H|^{2}=\{ |S|^{2}, |H_u|^{2}, |H_d|^{2}\}$ stands for the quadratics of 
the Higgs fields. These phase--dependent operators are all determined by the
$R$--invariance arguments. However, their contribution to the vacuum energy depends on 
the diagrammatics, and this will be done in the next section. The radiative stability of the vacuum
with respect to the phases will realize the relaxation process.

\begin{table}[htbp]
\begin{center}
\begin{tabular}{||c||c|c||}
$\mbox{Non-marginal operators (d = 2)}$&$\mbox{Marginal operators (d = 4)}$&$\mbox{phase content}$\\\hline \hline   
$A_{s}A_{k}^{*}$ & $A_{s}S^{3},\ \ A_{s}A_{k}^{*}|H|^{2}$&$\phi_{s}-\phi_{k}$\\\hline
$A_{f}A_{s}^{*}$ & $A_{f}S H_{u}\cdot H_{d},\ \ A_{f}A_{s}^{*}|H|^{2}$&$\phi_{f}-\phi_{s}$\\\hline
$A_{f}A_{k}^{*}$ & $A_{f}S^{3},\ \ A_{f}A_{k}^{*}|H|^{2}$&$\phi_{f}-\phi_{k}$\\\hline
$m_{\lambda}A_{s}^{*}$ & $m_{\lambda}S H_{u}\cdot H_{d}, \ \ m_{\lambda}A_{s}^{*}|H|^{2}$&$\phi_s$\\\hline
$m_{\lambda}A_{k}^{*}$ & $m_{\lambda} S^{3}, \ \ m_{\lambda}A_{k}^{*}|H|^{2}$&$\phi_k$\\\hline
$m_{\lambda}A_{f}^{*}$ & $m_{\lambda}A_{f}^{*}|H|^{2}$&$\phi_f$\\\hline
$\mbox{  }$&$m_{\lambda}A_{f}A_{k}^{*}A_{s}^{*}$&$\phi_{k}+\phi_{s}-\phi_{f}$\\\hline                        
$\mbox{  }$&$m_{\lambda}A_{f}^{*}A_{k}A_{s}^{*}$&$\phi_{f}+\phi_{s}-\phi_{k}$\\\hline
$\mbox{  }$&$m_{\lambda}A_{f}^{*}A_{k}^{*} A_{s}$&$\phi_{k}+\phi_{f}-\phi_{s}$
\end{tabular}
\end{center}
\caption{\label{table:back1}
Phase--dependent, $R$--invariant, $d=2$ and $d=4$ operators and their phases.}
\end{table}

\section{Dynamical Relaxation of the CP phases}
It is clear from the superpotential (\ref{nmssm}) that, the 
fermion superfields cannot be assigned $R$--charges like $R_{Q}=-R_{U}$  
or $R_{Q}=-R_{D}$ or $R_{L}=-R_{E}$; therefore, ${\cal{U}}(1)_{R}$ has
to have a quantum mechanical anomaly with respect to both QCD and QED. 
Then ${\cal{U}}(1)_{R}$  is a nonlinearly--realized global symmetry of the
lagrangian, and the phase fields above are nothing but the Goldstone bosons
of some spontaneously broken global symmetries in the SUSY--breaking sector.
The QCD anomaly of ${\cal{U}}(1)_{R}$ shifts the QCD vacuum angle as 
$\theta_{QCD} \longrightarrow \theta_{QCD} + {\cal{G}}_{R}(x)$. Since ${\cal{G}}_{R}(x)$
is a Goldstone boson it would have a strictly flat potential were not it
for the instanton effects in the QCD vacuum which develops a potential
such that $\langle {\cal{G}}_{R}(x) \rangle = - \theta_{QCD}$ whereby
solving the strong CP problem \cite{pq,ww,ksvz,dfsz}. The resulting
massive pseudoscalar is the $R$--axion, $a_R$, having mass 
$m_{R}\sim m_{\pi}f_{\pi}/M_{SUSY}$ and decay constant $f_R\sim M_{SUSY}$.
These axion parameters are in the invisible $axion$ $window$ \cite{axion} as long
as $M_{SUSY}$ refers to an intermediate scale. In fact, the $R$--axion here is
both a KSVZ axion (through the gluino phase) and DFSZ axion (through the 
phases of the Higgs doublets). Needless to say, exactly the same kind of 
relaxation effect occurs also in the MSSM with the associated 
$R$--symmetry \cite{relaxMSSM}.

The soft breaking lagrangian (\ref{pot}) consists of four independent phases 
$\mbox{Arg}[A_s]$, $\mbox{Arg}[A_k]$, $\mbox{Arg}[A_f]$ and $\mbox{Arg}[m_{\lambda}]$.
Clearly one combination of these four fundamental phases is spent in relaxing the QCD
vacuum angle. However, the remaining three phases $\phi_{s}(x)$, $\phi_{k}(x)$ and 
$\phi_{f}(x)$ cannot be determined through the QCD effects. Therefore, one must check
the long--distance (electroweak interactions) and short--distance (interactions 
around Planck scale) contributions to find their relaxation points.

Relaxation of the effective QCD vacuum angle occurs via its instanton--induced
potential. Similar to this, we now look for the possibility of inducing potentials
for the phases of the operators in Table I. To do this we compute the coefficients of 
these operators. The operators with mass dimension four (the second column in Table I) are 
$marginal$ operators, that is, their conribution to the vacuum energy is
always weighted by dimensionless coefficients. On the other hand, operators
with dimension two (first column of Table I) are $non$--$marginal$ operators
in the sense that their coefficients should have always mass dimension two.
The divergences of the $marginal$ operators could be at most $logarithmic$ 
whereas those of the $non$--$marginal$ ones are $quadratic$.

\begin{figure}[htd]
\begin{center}
\vspace{9pt}
\epsfig{file=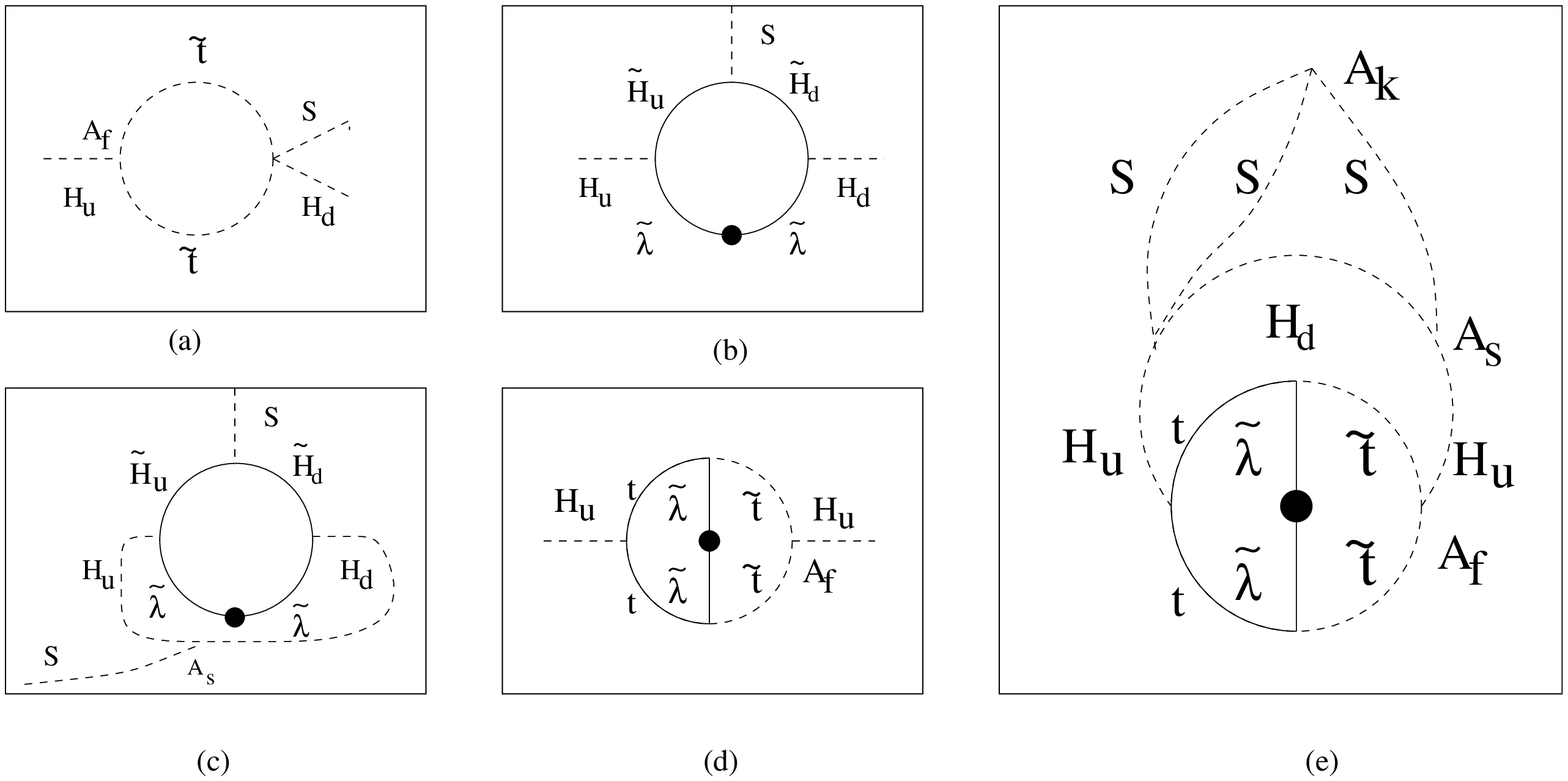, width=\linewidth}
\caption{Sample diagrams generating some of the $marginal$ operators in Table I. 
The dot on the gaugino line shows the gaugino mass insertion.}
\label{spect5}
\end{center}
\end{figure}

We first calculate the contributions of the $marginal$ operators to the vacuum energy. Depicted in Fig. 1 is 
a set of sample diagrams that generate some of the operators in the second column of Table I. Here (a), (b), (c)
and (d) generate, respectively, $h_{s}h_{f}^{\ast}A_{f} S H_{u}\cdot H_{d}$, $h_{s}m_{\lambda}S H_{u}\cdot H_{d}$,
$h_{s}m_{\lambda}A_{s}^{\ast}|S|^{2}$ and $h_{f} m_{\lambda} A_{f}^{\ast}|H_u|^{2}$. All the remaining dimension--four
operators can be generated using similar diagrams. For example, the diagram (e) in Fig. 1 generates $h_{t} k_{s}
h_{s}^{\ast}m_{\lambda}A_{f}^{\ast}A_{k}^{\ast} A_{s}$. One notices that evaluation of each diagram in Fig. 1 produces 
a $marginal$ operator belonging to Table I; however, each contribution is conveyed by a by a non--dynamical phase
associated with the Yukawa couplings. Letting the loop momenta vary from $m_{3/2}$ to $M_{Pl}$, on dimensional 
grounds, the $marginal$ operators contribute to the vacuum energy as
\begin{eqnarray}
\label{long}
\left(\Delta V\right)_{long}&=&m_{3/2}^{4} \log \left(\frac{M_{Pl}^{2}}{m_{3/2}^{2}}\right)\Bigg\{
c_{a} \cos \left( \phi_{s}-\phi_{f} + \theta_{s}-\theta_{f}\right)+ c_{b,c}\cos \left( \phi_{s} +
\theta_{s}\right)\nonumber\\
&+&c_{d} \cos \left( \phi_{f} + \theta_{f}\right)+c_{e} \cos \left( \phi_{f} +\phi_{k}-\phi_{s}+
\theta_{f}+\theta_{k}-\theta_{s}\right)+\cdots\Bigg\}
\end{eqnarray}
where the subscript $long$ emhasizes that this potential is sensitive to long--distance ($m_{3/2}^{-1}$)
physics; its dependence on short--distance ($M_{Pl}^{-1}$) physics is only logarithmic. In this expression the 
dimensionless parameters $c_{a},\cdots, c_{e}$ are, respectively, the weights of the diagrams $(a),\cdots,(e)$ in Fig. 1,
and the ellipses stands for contributions of the diagrams that generate other marginal operators listed in Table I.
Here the weight factors $c_{i}$ consist of the Yukawa and gauge couplings as well as the loop suppression factors.

After estimating the contributions of $marginal$ operators, we now compute those of the $non$--$marginal$ 
operators in the first column of Table I. Depicted in Fig. 2 are the loop diagrams ($(a),\cdots,(f)$) generating the 
relevant operators in rows ($1,\cdots,6$) of Table I, respectively. On dimensional grounds, the $non$--$marginal$ operators
contribute to the vacuum energy as follows
\begin{eqnarray}
\label{short}
(\Delta V)_{short}&=&\frac{M_{Pl}^{2}
m_{3/2}^{2}}{(4\pi)^{6}}\Bigg(\overline{c_{f}}\cos(\phi_{f}+\theta_{f}+\overline{\delta_{f}})+
\overline{c_{s}}\cos(\phi_{s}+\theta_{s}+\overline{\delta_{s}})\nonumber\\&+&
\overline{c_{sk}}\cos(\phi_{k}-\phi_{s}+\theta_{k}-\theta_{s}+\overline{\delta_{sk}})+
\overline{c_{fs}}\cos(\phi_{s}-\phi_{f}+\theta_{s}-\theta_{f}+\overline{\delta_{fs}})\nonumber\\&+&
\frac{\overline{c_{fk}}}{(4\pi)^{2}}\cos(\phi_{k}-\phi_{f}+\theta_{k}-\theta_{f}+\overline{\delta_{fk}})+   
\frac{\overline{c_{k}}}{(4\pi)^{2}}\cos(\phi_{k}+\theta_{k}+\overline{\delta_{k}})
\Bigg)
\end{eqnarray}
where the subscript $short$ stresses that this contribution is highly sensitive to short--distance
physics due to its quadratic dependence on $M_{Pl}$. In this expression, $\overline{c_{i}}$ stands for the weight 
of the $i$--th diagram in Fig. 2, and $\overline{\delta_{i}}$ its possible phase shift beyond the ones coming 
from the Yukawa couplings. One notices that $\overline{c_{i}}$ here does not include the loop factors;
they are functions of only Yukawa and gauge couplings. The additional phase shifts $\overline{\delta_{i}}$
represent possible sources of CP--violation at short--distances. If CP is spontaneously broken together
with supersymmetry they can be taken at the CP conserving points. This is the case is in the supergravity
scenarios \cite{stringCP} so that we will take it granted by setting all $\overline{\delta_{i}}$ to zero \cite{relaxMSSM}.
 
\begin{figure}[htd]
\begin{center}
\vspace{9pt}
\epsfig{file=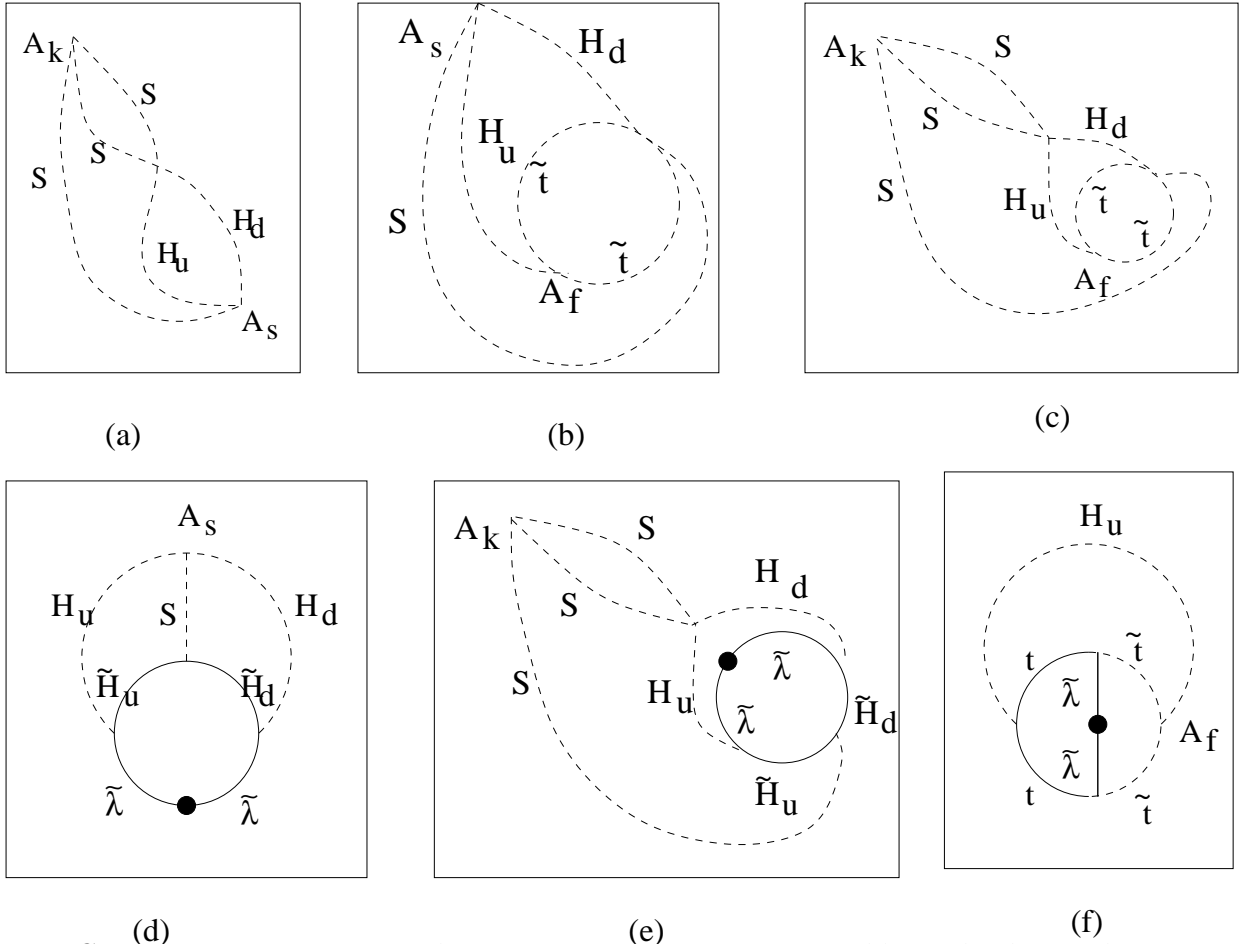 ,width= \linewidth}
\caption{Diagrams generating the $non$--$marginal$ operators in Table I. The dot on the gaugino line 
represents the gaugino mass insertion.}
\label{spect9}  
\end{center} 
\end{figure}

So far we have assumed all Yukawa couplings to have arbitrary nondynamical phases. However, it is possible 
to rotate the quark fields such that the phases of $h_u$ and $h_d$ are transferred to charged current vertices
via the CKM matrix. In the SM the CKM matrix is the only source of CP violation, and its size is 
characterized by the Jarlskog invariant $J={\cal{I}}m[V_{ud}V_{td}^{\ast}V_{td} V_{ub}^{\ast}]$.
Contribution of this parameter to the vacuum energy is of ${\cal{O}}(\alpha/4\pi)^{2} J)$ which is much
smaller than the supersymmetric ones. Indeed, the Higgs sector Yukawa couplings ($h_s$ and $h_k$)
ahve important differences from $h_u$ and $h_d$ concerning the loop structures in Figs. 1 and 2.

The vacuum energy with radiative corrections takes the form 
\begin{eqnarray}
\label{totpot}
V(\phi_{f},\phi_{s},\phi_{k})&=&V_{tree}+(\Delta
V)_{short}(\phi_{f},\phi_{s},\phi_{k})+(\Delta V)_{long}(\phi_{f},\phi_{s},\phi_{k})
\end{eqnarray}
which has to be minimized with respect to $\phi_{k}(x)$, $\phi_{f}(x)$ and $\phi_{s}(x)$. It is worth noting 
that there is no particular operator in Table I which has only long--distance sensitivity, that is, all 
of the dynamical phases appear in both (\ref{long}) and (\ref{short}). One further notes that $V_{short}$ 
has a large weight factor due to its quadratic $M_{Pl}$ dependence. Therefore, all extremization equations 
for (\ref{totpot}) are saturated by the short--distance contribution, 
\begin{eqnarray}
\label{vev}
\langle\phi_{f}\rangle=\langle\phi_{k}\rangle+\theta_{k}=\langle \phi_{s}\rangle+\theta_{s}+\pi
\end{eqnarray}
after assuming that all $\overline{c_{i}}$ are of similar order of magnitude, which is reasonable. 
It is worthy of noting that none of the phases develops a VEV in close proximity of a CP--conserving point: $\langle\phi_{f}\rangle,
\langle\phi_{k}\rangle, \langle \phi_{s}\rangle \neq 0, \pi$. It is mainly here that there is an important difference between the 
NMSSM and the MSSM: The latter has all phases relaxing the CP--conserving points whereas the former does not.
In the next section we will discuss the reason for this difference by considering the MSSM limit of the NMSSM.

The physical Goldstone bosons ${\cal{G}}_{f,k,s}(x)\equiv M_{SUSY}~\left(\phi_{f,k,s}(x)-\langle\phi_{f,k,s}\rangle\right)$, 
are massive pseudo--scalars with masses
\begin{eqnarray}
\label{massfin}
m_{k, s, f}^{2}\sim M_{Pl}^{2} m_{3/2}^{2}/M_{SUSY}^{2}.
\end{eqnarray}
Therefore, particular short--distance sensitivity of the potentials of $\phi_{f,k,s}(x)$ require the pseudo--Goldstone
bosons ${\cal{G}}_{f,k,s}(x)$ to have masses right at the intermediate scale. These pseudo--Goldstone bosons have only 
derivative couplings to the visible matter so that they are invisible to collider experiments.

During the entire analysis the triscalar couplings in (\ref{pot}) are written without  Yukawa couplings, for convention.
If required, one can seperate the Yukawa couplings by the replacement, $A_{s}\rightarrow h_{s} A_{s}$, $A_{k}\rightarrow k_{s}
A_{k}$ and $A_{f}\rightarrow h_{f} A_{f}$. Then the VEV's in (\ref{vev}) give
\begin{eqnarray}
\label{baska}
\Big<\mbox{Arg}[m_{\lambda} A_{f}^{\ast}]\Big>=\Big<\mbox{Arg}[m_{\lambda} A_{k}^{\ast}]\Big>=\Big<\mbox{Arg}[m_{\lambda}
A_{s}^{\ast}]\Big> +\pi~,
\end{eqnarray}
which are independent of the Yukawa phases. This replacement seperates dynamical and nondynamical phases, 
and still VEV's of the phases do not relax to a CP--conserving points. From this relation it follows that 
if any of these phases relaxes to a CP--conserving point by some reason so do the remaining two. 

\section{The MSSM limit}   
To clarify the meaning of the CP--violating relaxation points in (\ref{vev}) or (\ref{baska}) it may be convenient to discuss
the MSSM limit both algebraically and diagrammatically. The main difference between the MSSM and NMSSM follows from their
symmetries and structures of the superpotentials. With purely triscalar nature of the soft terms in (\ref{pot}), the scalar
potentials of all pseudo--Goldstone bosons turn out to be controlled by the short--distance physics, in particular, there is no
phase--dependent invariant that receives a potential only from the long--distance physics. This is not the case in the MSSM as
one of the phases receives a potential only from the long--distance effects \cite{relaxMSSM} so that its relaxation dynamics is
different than those of the remaining two. In the conventions leading to (\ref{baska}), the MSSM limit is realized by the
replacements
\begin{eqnarray}
\label{limit}
h_{s}S\rightarrow \mu,\ \ \ h_{s} A_{s} S \rightarrow m_{12}^{2},\ \ \ k_{s}\rightarrow 0,
\end{eqnarray}
where now  $\mu$ and $m_{12}^{2}$ are no longer dynamical fields, instead they are background spurions that
can appear only as mass insertions in the loop diagrams. It may be convenient to discuss the modifications in the topologies of the
diagrams in Fig. 2 under the replacements (\ref{limit}). It is clear that diagrams (a), (c) and (e) vanish due to vanishing
$k_{s}$. However, as Fig. 3 shows explicitely, the three--loop diagrams (b) and (d) in Fig. 2 go over to two--loop diagrams
($\overline{b}$), ($\overline{d}$)  whereas the diagram (f) remains unaffected. A simple observation on Fig. 3 shows that
($\overline{b}$), ($\overline{d}$) and ($\overline{f}$) generate, respectively, the phases $\mbox{Arg}[\mu A_{f}
{m_{12}^{*}}^{2}]\equiv \phi_{s}-\phi_{f}$, $\mbox{Arg}[m_{\lambda} \mu {m_{12}^{*}}^{2}]\equiv \phi_{s}$ and
$\mbox{Arg}[m_{\lambda} A_{f}^{*} ]\equiv \phi_{f}$. It should be emphasized that these phases are precisely the ones 
appearing in the MSSM \cite{relaxMSSM}.
 
\begin{figure}[htd]
\begin{center}
\vspace{9pt}
\epsfig{file=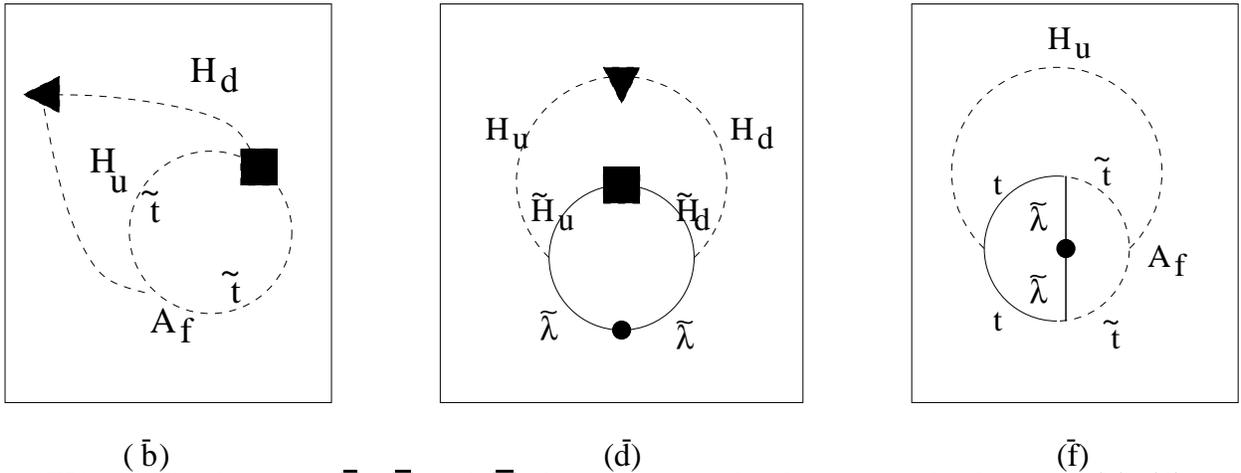 ,width= \linewidth}
\caption{The diagrams ($\overline{b}$), ($\overline{d}$) and ($\overline{f}$) show, respectively, the form of the diagrams
(b), (d) and (f) of Fig. 2 in the MSSM limit. Here the dot, triangle and square correspond to the insertions of the mass
parameters, $m_{\lambda}$, $m_{12}^{2}$ and  $\mu$, respectively.}
\label{spect10}  
\end{center}  
\end{figure}

As is clear from the modifications in the topologies of the diagrams in Fig. 2 and Fig. 3, types of the divergences 
of the diagrams are modified. Indeed, unlike (b) and (d) of Fig. 2, now ($\overline{b}$) and ($\overline{d}$) are only
$logarithmically$ divergent. This is, in fact, what is implied by the spurion character of the singlet field in the MSSM limit. It is
this effect that generates the $marginal$ operators $\mu A_{f} {m_{12}^{*}}^{2}$ and  $m_{\lambda} \mu {m_{12}^{*}}^{2}$ which are
only logarithmically divergent. On the other hand, the MSSM limit does not alter the logarithmic structure of (\ref{long})
apart from certain modifications in individual diagrams. Therefore, summing up the contributions of all diagrams, the radiative
corrections to the vacuum energy takes the form 
\begin{eqnarray}
(\Delta V)_{\mbox{\small MSSM}}&=&m_{3/2}^{4} \log \left(\frac{M_{Pl}^{2}}{m_{3/2}^{2}}\right)\Bigg[ c_{s} \cos (\phi_{s}
+\delta_{s})
 +c_{sf} \cos(\phi_{s}-\phi_{f}+\delta_{sf})\Bigg]\nonumber\\
&+&m_{3/2}^{2} M_{Pl}^{2} c_{f} \cos (\phi_{f} +\delta_{f})~.
\end{eqnarray}
In this expression the weight factors and the phase shifts have the same meaning as in (\ref{long}) and (\ref{short}), and the
loop suppression factors are not factored out. Taking again the phase shifts at CP--conserving
points in both short-- and long--distance contributions, one finds that $\langle\phi_{f}\rangle =0$ or $\pi$ to an accuracy 
${\cal{O}}(m_{3/2}^{2}/ M_{Pl}^{2})$. Then, necessarily $\langle \phi_{s}\rangle=0$ or $\pi$. Therefore, $\langle
\phi_{f}\rangle$
($\langle\phi_{s}\rangle$) is determined solely by the short--distance  (long--distance) dynamics. Moreover, both phases relax to
CP--conserving points. As a result, in the MSSM limit (\ref{limit}) the phases in (\ref{vev}) and (\ref{baska}) go over the usual 
MSSM relaxation pattern leaving $\phi_{k}$  completely undetermined. 

It is, in fact, the  $k_{s}\rightarrow 0$ limit that washes out any information about the fate of $\phi_{k}$. 
Obviously, for $k_s\equiv 0$, the NMSSM superpotential (\ref{nmssm}) possesses an additional $U(1)_{PQ}$ symmetry, and therefore, 
$\langle\phi_k\rangle$ is nothing but the phase lifted by this global symmetry. Therefore, relaxation of the physical CP
violating phases away from the CP--conserving points in (\ref{vev}) or (\ref{baska}) follows from the fact that this $U(1)_{PQ}$
symmetry is broken by a dimensionless parameter $k_{s}$ carrying no information about the SUSY breaking sector. 

\section{Conclusions and Discussions}

In Table II we list the main results of the work in comparison with the MSSM predictions \cite{relaxMSSM}. Here the 
upper block refers to the non--dynamical phases whereas the lower block is for the same models with dynamical phases.
As the table shows, all three NMSSM pseudo--Goldstone bosons have masses around $M_{SUSY}$ so that they are too 
heavy to appear in the existing colliders. On the other hand in the MSSM one of the pseudo--Goldstone bosons fall below
the TeV--scale.
 
\begin{table}[htbp]
\begin{center}
\begin{tabular}{||c||c|c||}
$\mbox{ }$&$\;\;\;\;\;\;\;\;\;\;\;\;\;\;\;\;\;\;\;\mbox{MSSM}$&$\mbox{NMSSM}$\\\hline\hline
$\mu-\mbox{problem}$&$\mbox{yes}$&$\mbox{no}$\\\hline
$\mbox{domain walls}$&$\mbox{no}$&$\mbox{yes}$\\\hline
$\mbox{strong CP problem}$&$\mbox{yes}$&$\mbox{yes}$\\\hline
$\mbox{sources for SUSY CP violation}$&$\mbox{yes}$&$\mbox{yes}$\\\hline
$\mbox{global symmetries}$&$\mbox{ }$&${\cal{Z}}_{3}$
\end{tabular}
\vspace{0.3cm}
\begin{tabular}{||c||c|c||}
$\mu-\mbox{problem}$&$\mbox{yes}$&$\mbox{no}$\\\hline
$\mbox{axionic domain walls}$&$\mbox{yes}$&$\mbox{yes}$\\\hline
$\mbox{strong CP problem}$&$\mbox{no}$&$\mbox{no}$\\\hline
$\mbox{sources for SUSY CP violation}$&$\mbox{no}$&$\mbox{yes}$\\\hline\hline
$\#~\mbox{of pseudo--Goldstone bosons}$&$2$&$3$\\\hline
$\mbox{and their masses}$&$\sim m_{3/2}^{2}/M_{SUSY}$ and $\sim M_{SUSY}$&$\mbox{all have}~\sim M_{SUSY}$\\\hline\hline
$\mbox{global symmetries}$&$U(1)_{PQ}\times U(1)_{R}$&${\cal{U}}(1)_{R}$
\end{tabular}
\end{center}
\caption{\label{table:back2}
MSSM vs NMSSM without (uper block) and with (lower block) dynamical phases. Unlike
the MSSM, the NMSSM provides sources for SUSY CP violation, solves the $\mu$ problem, and 
has all its physical Goldstone bosons with masses ${\cal{O}}(M_{SUSY})$. Moreover, both models 
solve the strong CP problem at the expense of developing axionic domain walls.}
\end{table}

Having finite SUSY CP violation with vanishing QCD vacuum angle is an important property of NMSSM 
not found in the MSSM. Indeed, with finite $\theta_{QCD}$, it does not matter if one cancels the SUSY contributions to EDMs
\cite{cancel}, because the QCD contribution anyhow exceeds the bound by several orders of magnitude. Therefore, it is in the NMSSM
that one can safely take SUSY and CKM phases as the mere sources of CP violation, that is,  EDMs of the electron and neutron as
well as $\epsilon_K$ and  $\epsilon_K^{\prime}/\epsilon_{K}$ can all be calculated without worrying about the effects of 
the nonperturbative strong interactions. It is a question of the SUSY parameter space if the calculated values for these 
observables agree with the experiment.

It may be useful to recall here the relevance of the flavor structure. Given the CLEO determination
of BR$(B\rightarrow K^\ast \gamma)$ then there is no possibility that the SUSY CP violation can saturate the observed 
CP violation with minimal flavor structure \cite{biz}. Though having finite SUSY CP violation is necessary it is by
no means sufficient; one has to look for flavor structures beyond the usual Yukawa hierarchies.

It is a rather general statement \cite{sikivie} that all Peccei--Quinn type solutions to the strong CP problem suffer
from the axionic domain walls, which are diastrous cosmologically \cite{tamvakis,zko}. Altough axionic wall formation is an additional
problem for the MSSM, it exists in the NMSSM with and without the dynamical phases so that phase relaxation does not generate a
new difficulty in this model. One of the ways of avoiding the axionic wall formation is to embedd the $R$--symmetry into
the center of a GUT group \cite{embed}. However, in non--GUT gauge structures like MSSM and NMSSM this method does not work.
The only remaining way out of this difficulty is to choose nonminimal flavour structures \cite{horizon} such that domain wall 
number is reduced to unity. It is in this sense that the question of having observable SUSY CP violation and avoiding the 
domain walls might be answered by a common flavor structure. One further notes that the quadratic short--distance sensitivity
of the NMSSM vacuum energy arises also in the dynamical flavor matrices \cite{flavor}, Yukawa couplings \cite{yukawa} and
determinations of the SUSY--breaking  scale \cite{susyscale}. 

\section{Acknowledgements} 
The author is grateful to Goran Senjanovi\'c for numerous discussions and helpful comments about  
various aspects of this work. The author thanks to Charanjit  Aulakh and Antonio Masiero for useful discussions 
and their careful reading of the manuscript. 

\end{document}